**Application of the theory of Linear Singular Integral Equations and Contour Integration to Riemann Hilbert Problems for determination of new decoupled expressions of Chandrasekhar's X- and Y- functions for slab geometry in Radiative Transfer.**

**Rabindra Nath Das***

Visiting Faculty Member , Heritage Institute of Technology , Chowbaga Road, Anandapur , P.O.East Kolkata Township, Kolkata-7000107, West Bengal , India ,  Email address : rndas1946@yahoo.com

**Abstract:**  In Radiative transfer, the intensities  of radiation  from the bounding faces of finite slab are obtained in terms of X- and Y- functions of Chandrasekhar . Those  are non linear non homogeneous coupled integral equations . Those non linear integral equations are meromorphically extended to the complex plane to get linear non homogeneous coupled integral equations. Those linear integral equations are converted to linear singular integral equations with. linear constraints . Those singular integral equations are then transformed to non homogeneous Riemann Hilbert Problems. Solutions of those Riemann Hilbert Problems are obtained using the theory of linear singular integral equations to decouple those X- and Y- functions. New forms of linear non  homogeneous decoupled integral equations are derived for X- and Y- function separately with new  linear constraints. Those new decoupled integral equations are transformed into linear singular integral equations to get two new separate non homogeneous Riemann Hilbert problems  and  to find solutions in terms of one known N- function and five new unknown functions in complex  plane . All five functions are represented in terms of  N-functions using the theory of contour integration . Those X- and Y- functions are finally expressed  in terms of that N - function and  also in terms of H- functions of Chandrasekhar   and of integrals in Cauchy principal value sense in the complex plane and real plane. both for conservative and non conservative cases . The   H − functions for semi infinite atmosphere are derived as a limiting case  from the expression  of X- function of finite atmosphere.

 ***Permanent Address:  KB −9 , Flat − 7 , Sector − III , Bidhan nagar  ( Salt Lake ) , Kolkata- 700098 , West Bengal , India .

.

**Introduction :** Most of the completely solved problems in radiation transport theory are concerned with atmosphere stratified in plane parallel layers . Every problem is essentially that of solving the " equation of transfer " ( an integro differential equation ) subject to given boundary conditions .By certain standard techniques , the problems can be reduced to the solutions of non linear integral equations . This reduction is for determination of intensity of radiation from the bounding faces of atmosphere and also for the intensity of radiation at any optical depth of the atmosphere. In semi infinite atmosphere the intensity of radiation can be determined when the solution of the non linear integral equation for H- functions of Chandrasekhar[1] is obtained in closest form in terms of known functions. In case of finite atmosphere, the intensity of radiation at the boundaries and at any depth are obtained when explicit solution for the coupled integral equations of the X- and Y- functions of Chandrasekhar[1] are known. Two cases of those problems are also of prime practical importance. The problems of conservative case are that when there is no true absorption of radiation and the efficiency of scattering is unity ( solar atmosphere); the problems of non conservative cases are that when on each scattering , more radiation was not emitted than was incident ( planetary atmosphere) .

Chandrasekhar [1] solved the problems of diffuse reflection and transmission of light scattering in his study of Radiative transfer in homogeneous plane parallel planetary atmosphere of finite thickness by the principle of invariance . He introduced two non linear and non homogeneous coupled integral equations for X – and Y- functions for determining the intensity of radiation at the boundaries of the finite atmosphere and for determination of scattering functions and transmission functions .He did not establish the existence and uniqueness of solutions of those functions .

Busbridge [2] used the Ambartsumian technique combined with the theory of N- solutions of an auxiliary equation and assured the existence of X- and Y- functions in real and complex plane .She outlined analytic properties for those X- and Y – functions in the complex plane. She showed that every solution of her linear integral equations for X and Y- functions were not solution of non linear coupled integral equations of Chandrasekhar[1].

Chandrasekhar and Elbert [3] obtained numerical solutions of coupled non linear integral equations of Chandrasekhar[1] by iteration .

Busbridge [4] transformed her linear integral equations of X- and Y- functions to linear singular integral equations and obtained Fredholm integral equations for those and she proved that her Fredholm operators were contracting for sufficiently large atmosphere

Brandt and Chamberlain [5] obtained iterative solutions of coupled non linear integral of X- and Y- functions only for isotropic scattering for small values of optical thickness of the atmosphere and only for albedo for single scattering equal to unity

Mullikin [6] gave an exact criterion for uniqueness of solutions to non linear X- and Y- equations .In case of non uniqueness in non conservative cases , he gave a simple representation of all solutions in terms of the solution of the auxiliary equation of Busbridge [2].

Mayers [7] was able to obtain accurate solutions to non linear coupled integral equations of X- and Y- equations combining the method of Chandrasekhar and Elbert [3] and a method based on iterative solution of Schwarzschild-Milne integral equation .

Sobouti [8] tabulated the X- and Y- functions for different values of optical depth .

Mullikin [9] completed the linear coupled integral equations of X- and Y- functions by addition of linear constraints for those functions ..He first used the theory of linear singular integral equations of Muskhelishvilli [10] to obtain the linear coupled singular integral equations for X- and Y- functions and linear constraints. He also specified a unique pair of functions with the requirement of analyticity in the half complex plane for relating X- and Y – functions in terms of those . He used analytic continuation to transform their first set of Fredholm equations to different Fredholm equations with simpler continuous and non negative kernels . He obtained Fredholm equations for values of the X- and Y- functions in an interval on real axis of complex plane which were suitable for numerical computation He showed that the new Fredholm operators are contracting for all values of the atmospheric thickness with very rapid convergence for thick atmosphere .He obtained analytic expression X- and Y- functions in the complex plane in terms of H- functions of Chandrasekhar[1] and of some other function but those are dependent on the coupled X-and Y- functions. He obtained an expression of H- function also.

Carlstedtt and Mullikin [11] also solved numerically and prepared extensive tables of X- and Y- functions for different values of albedo for scattering .

Bellman , Kagiwaga and Ueno [12] presented graphs and selected tables of numerical values of those X and Y functions covering wide range of slab thickness and albedos for single scattering .

Adams and Kattawar [13] obtained scattering and transmission matrices according to Rayleigh Phase Matrix.

Fymat and Abhyankar [14] derived the non linear integral equations in X-,Y-, X*- and Y*- functions for inhomogeneous atmospheres similar to thee Chandrasekhar's equations for homogeneous atmospheres . The relations to Ueno's and Chandrasekhar's equations are discussed and two approaches to the treatment of the problems of uniqueness of the solutions of these equations are suggested

Rybicki [15] in the search light problem with isotropic scattering. used the non linear integral equations for H , X and Y and derived numerical results by means of kernel approximation methods.

Caldwell [16] described a non iterative method for solving the X- and Y-functions and obtained tables of values of those functions and their moments . He used the coupled integro- differential equations satisfied by the X- and Y- functions of Chandrasekhar [1] in stead of non linear coupled integral equations for X- and Y- functions with Lobatto quadrature formulae to solve 96 coupled equations .

Bellman, Poon ,and Ueno [17] used coupled integral recurrence relation for finite order only for conservative isotropic scattering.

Domke [18] framed singular integral equations and corresponding linear constraints for several standard problems of radiative transfer in semi infinite and finite atmosphere. The stokes vector of emergent radiation obey independent half range singular integral equations where as the inner radiation field at any optical depth can be found from a full range integral equation depending on the radiation field at the boundary.

Das[19,20],Das [21,22],Das [23.24],Das[25]Das[26] developed an exact method of Laplace Transform in combination with the theory of linear singular integral equations and solved the integro-differential equations arising in the study of radiative transfer for finite steady , plane parallel , isotropic and anisotropic scattering atmosphere for intensity of radiation from the bounding faces and also at any depth of the atmosphere in terms of X- and Y- functions. He outlined that the origin of the linear integral equations satisfied the X- and Y- functions of Chandrasekhar[1] was within the integro differential equation of radiative transfer if one applies the Laplace transform to that basic Integro -differential equation..He used the Wiener- Hopf technique to determine the exact form of X- and Y- functions in terms of H- functions .His exact form was also coupled and interdependent .

Dasgupta[27] ; Dasgupta and Bishnu [28] outlined the properties of exponential functions in representing the X- and Y- functions .

Haggag , Machali and Madkour [29] used the Galarkin method to calculate those X- and Y- functions and their moments.

Lahoz [30, 31] presented an exact method for linearzing and decoupling the nonlinear coupled integral equations of X- and Y- functions of Chandrasekhar[1].He introduced two new functions: the first one satisfies a non- linear singular integral equation and the second is a function integrally dependent on first function . Those decoupled linear singular integral equations for X- and Y- functions are in principle solvable analytically using the theory of Linear singular equations of Muskhelishvilli [10]. He obtained X- and Y- functions as an arbitrary solution of linear decoupled singular integral equations in terms of his two new functions . Arbitrariness remained in his solution but his solutions were decoupled .

Biswas and Karanjai [32] derived time dependent H-. X- and Y- functions for an anisotropically scattering atmosphere.. The method of integral operator of Ambartsumian and theory of N- solution of Busbridge have been used to find the corresponding H- function and X- and Y- functions.

Roy Mondal , Biswas and Karanjai [33] derived frequency dependent X- and Y- functions for a non- coherent scattering problems by the combination of eigen function approach of Case [34] and from the principle of invariance of Chandrasekhar [1].

Danielian [35,36]used the methods of separated linear integral equations and obtained a complete analytic solution which are interdependent .

Rutily ,Chevallier and Bergeat [37,38]  dealt with the singular integral equations satisfied by the  X- and Y -equations  in Cauchy principal value sense. They obtained X-and Y- function represented in terms of two new functions . Their new functions satisfy Fredholm  integral equations.

Das [39,40] obtained new representations of H- functions of Chandrasekhar[1]using  Wiener- Hopf technique and theory of singular integral equations to Riemann Hilbert problems .

In this paper we considered two  non linear non homogeneous coupled integral equations for X- and Y- functions in the complex plane with linear constraints . Those are converted to linear non homogeneous coupled singular integral equations in real plane using Plemelj's formulae  .We obtained two coupled Riemann Hilbert Problems for solving  X - and Y - functions  using the theory of linear singular integral equations  .Using the method outlined by Das [40] those coupled  linear Hilbert problems for X- and Y- functions are converted two new linear non homogeneous decoupled  integral equations for X and Y- functions in the complex plane . Those new decoupled integral equations are dependent on two new  unknown  $N_1$- and $N_2$- functions  and  one known  N- function of Mullikin [9] whose exact closest form is available in  Das [39,40]. We again transformed those two new linear non homogeneous decoupled singular  integral equations for X and Y- functions using Plemelj"s formulae. .We again converted those two  new decoupled linear singular integral equations to two new Riemann Hilbert Problems for each X- and Y- function . We again used the method outlined by Das [40] to  express  those decoupled X- and Y- functions exclusively and exactly   in terms H- functions of Chandrasekhar[1] and/ or in terms  of known N- function and in terms of five new unknown functions . Those five unknown functions are expressed in terms of known N- function using the theory of contour  integration . Those analytic results are new from theoretical stand point and also numerical point of view. All exact representation of X- and Y- functions so far obtained   by different authors are either coupled and interlinked or dependent of some other unknown functions. Here our method  of expressing X – and Y- function in both conservative and non conservative cases of finite atmosphere in terms of H- function and/ or in terms  of only  N- function  is completely new and  free from the interdependence . The H- function  is also only dependent on  known N - function whose form is explicitly known in Das[39] and Das[40]  .We also derived  the Chandrasekhar's  H- function  as a limiting case of X- function in semi-infinite  plane  parallel  atmosphere  both  for  conservative   and  non conservative cases .The unknown constants are derived from new constraints .

### I. Mathematical   formulation :

In plane parallel finite atmosphere, Chandrasekhar[1] derived   the X-Y functions . Those X-Y functions satisfy two    nonlinear non homogeneous coupled  integral equations  of the form :

$$X(u) = 1 + u \int_0^1 U(x) [ X(x) X(u) - Y(x) Y(u) ] \, dx / (x + u), \quad 0 \le u \le 1, \quad (1)$$

$$Y(u) = \exp(-t/u) + u \int_0^1 U(x) [Y(x) X(u) - X(x) Y(u)] \, dx / (x - u), \quad 0 \le u \le 1 \quad (2)$$

Here t is the optical thickness of the atmosphere. U (x) ,in general an even function called characteristic function  satisfies   the condition :

$$U_0 = \int_0^1 U(x) \, dx \le 1/2 \quad . \tag{3}$$

 The case $U_0 = 1/2$ is called the conservative case and the case $U_0 < 1/2$  is called the non-conservative case
The functions X(u) and Y(u) are related by the equation

$$Y(u) = \exp(-t/u) \quad X(-u) \tag{4}$$

In semi infinite atmosphere. the equations (1) and (2) play the similar role of H –function  and we require

$$X(u) \to H(u) \qquad \text{as } t \to \infty \, , \tag{5}$$

$$Y(u) \to 0 \qquad \text{as } t \to \infty \, . \tag{6}$$

where H(u) satisfies the non linear non homogeneous integral equation of the form

$$H(x) = 1 + x \, H(x) \int_0^1 U(t) \, H(t) \, dt / ( t + x) \quad , \qquad 0 \le x \le 1 \, , \tag{7}$$

More over, it will also appear that

$$X(u) \to 1 \text{ as } \quad t \to 0 \, , \tag{8}$$

$$Y(u) \rightarrow \exp(-t/u) \text{ as } t \rightarrow 0 \ . \qquad (9)$$

We also require that

$$X(u) \rightarrow 1 \text{ as } u \rightarrow 0 \ , \text{ from within } (0,1) \qquad (10)$$

$$Y(u) \rightarrow 0 \text{ as } u \rightarrow 0 \text{ from within } (0,1). \qquad (11)$$

Following Busbridge [2] equations.(1) and (2) can be transformed into linear non homogeneous coupled integral equations[LNCIE] in the complex z plane cut along (-1,1) to the following forms :
.

$$X(z) \ T(z) = [1 - z \exp(-t/z) \int_0^1 Y(t) \ U(t) \ d \ t / (t+z) \ ]$$

$$+ \ z \ \int_0^1 X(t) \ U(t) \ d \ t / (t-z) \qquad (12)$$

$$Y(z) \ T(z) = \exp(-t/z) \ [ \ 1 - z \ \int_0^1 X(t) \ U(t) \ d \ t / (t+z) \ ]$$

$$+ z \int_0^1 Y(t) \ U(t) \ d \ t / (t-z) \ . \qquad . \qquad (13)$$

where

$$T(z) = \ 1 - z \ \int_0^1 \ U(t) \ [ \ 1/(t+z) - 1/(t-z) \ ] \ d \ t \qquad (14)$$

$$= \ [ \ 1 - z \ \int_0^1 \ X(t) \ U(t) \ d \ t/ (t+z) \ ] \ [ \ 1 + z \ \int_0^1 \ X(t) \ U(t) \ d \ t / (t-z)]$$

$$+ z^2 \ [ \ \int_0^1 \ U(t) \ Y(t) \ d \ t / (t+z) \ ] \ [ \ \int_0^1 \ Y(t) \ U(t) \ d \ t / (t-z) \ ]. \qquad (15)$$

The function $T(z)$ called dispersion function , has the following properties :

i) $T(z)$ is analytic in the complex z plane cut along (-1,1),
ii) it has two logarithmic branch points at $-1$ and $+1$ ,
iii)$T(z) \rightarrow 1$ as $| z | \rightarrow 0$ ,
iv) it has only two simple zeros at infinity when $U_0 = \frac{1}{2}$ ,
v) only two real simple zeros at $z = +1/k$ and $z = -1/k$
   where k is real $0 < k \leq 1$ when $U_0 < \frac{1}{2}$ ,
vi) $T(z) \rightarrow -C/z^2$ as $z \rightarrow \infty$ when $U_0 = \frac{1}{2}$, $C = 2 \ U_2$,
    a real positive constant where $U_r = \int_0^1 u^r U(u) \ d \ u, \ r = 1,2,3....$
vii ) $T(z) \rightarrow D = 1 - 2 \ U_0$ as $z \rightarrow \infty$ when $U_0 < 1/2$ ,
    D is a real positive constant ,

viii)$T(z)$ is   bounded on the entire imaginary axis ,

ix) $T(\infty) = D = 1 - 2\,U_0$

$$= (\,1 - \int_0^1 H(x)\,U(x)\,dx\ )^2$$

$$= 1/\ (\ H(\infty)\ )^2,\ \text{when}\ \ U_0 < 1/2\ ,$$

x) it is an even function of z ,

xi) $T(z)$   is assumed to be negative  as $z \to 1+0$   when  $U_0 < \frac{1}{2}$ ,

xii) on assumption of Holder   condition  on the function  $U(u)$

in the interval  (0,1), $T(z)$ becomes   singular and takes the form

$$T_c(x) = 1 - 2\,x^2\ \int_0^1\ (\,U(u) - U(x)\ )\,d\,u\,/\,(x^2 - u^2)$$

$$-\,x\,U(x)\ \ln\ (\,(\,1+x)\,/(1\text{-}x)\,)\,),\qquad\qquad\qquad (16)$$

Busbridge [2] observed that every solution of equations. (12) & (13) is not a solution of equations (1) & (2) .She proved that in conservative case ( $U_0 = 1/2$) the solutions of equations(12) and (13) which are regular for l z l > 0  and O (z) as $z \to \infty$ and tends to a finite limit as l z l $\to$0 in the sector   $Z_1 =$   l arg(z) l $\le \beta < \pi/2$ are also the solution of equations (1) and (2). She also observed that in non conservative case ( $U_0 < 1/2$) , her all solutions  of equations (12) & (13) are not solutions of equations(1) & (2).

The equations ( 12) and (13) can be rewritten in the form :

$$X(z)\ T(z) = f_{Xc}(z,t) + z \int_0^1\ X(x)\ \ U(x)\,d\,x\,/\,(x\ -\,z)\qquad\qquad (17)$$

$$Y(z)\ \ T(z) = f_{Yc}\,(z,t)\ + z \int_0^1\ Y\,(x)\,U(x)\ \ dx\,/\,(\,x - z\,)\qquad\qquad (18)$$

where

$$f_{Xc}(z,t)\ =\ \ [1 - z\,\exp(\text{-}t/z)\int_0^1\ \ Y(t)\,U(t)\,d\,t\,/\,(\,t + z)\ \ ]\qquad\qquad (19)$$

$$f_{Yc}(z,t)\ =\ \ \exp(\text{-}t/z)\ [\ 1 - z\ \int_0^1\ \ X(t)\,U(t)\,d\,t\,/\,(t + z)\ ]\qquad\qquad (20)$$

where  $f_{Xc}(z,\ t)$   and   $f_{Yc}(z,\ t)$  coupled with $Y(z)$ and $X(z)$ respectively are continuous functions across the cut along (0,1), analytic in the right half plane Re(z) > 0  in the sector  $Z_1 =$  l arg(z) l $\le \beta < \pi/2$  .

The function

$$h(t\ ,z\ ) = \exp(\text{-}t/z)\qquad\qquad\qquad\qquad\qquad (21)$$

requires special mention .Following, Busbridge[2], Dasgupta [27 ] , Dasgupta and Bishnu [28]. Lahoz [30,31] it can be stated that

i)      $h(t,z)$ is an increasing function of $z=x+iy$ for Re $z >0$ ,

ii)       it has a point of inflexion at $z= t /2$,

iii)       $h(t,z) < x \exp(-t)$ for $x$ in $[0 , 1]$ and $1 \le t$ ,

iv)      $h(t,z) < x ( t \exp(-t) ) + \exp(-t) (1-t)$ for $x$ in $[0,1]$ and $t<1$.

v)      $O(\exp(-\alpha\ t/z) \to 0$ as $z \to 0$ in sector $Z_2= [ z; \text{Re } z >0, 0<\alpha \le 1, t>0]$,

vi)      Is bounded at the origin when $z \to 0$ along the imaginary axis,

vii)      $\to 1$ as $z \to \infty$,

viii)      $\to$ a constant  as $z \to 1$,

ix)      $\to \infty$ as $z \to 0$ with in sector $Z_3 = [ z ; \text{Re } z <0 , t >0 ]$,

has  an essential singularity  a t  $z \to 0-$,

x)      $| h(t,z) | = \exp( -t\ r^{-1} \cos\theta ) \le 1$ when $z = r \exp(i\theta )$ , $| \theta | \le \pi/2$ ,

xi)      $| h(t,z) | \le \exp( -t\ r^{-1} \cos\beta )$  when $z = r \exp(i\theta )$ , $| \theta | \le \beta <\pi/2$ .

Mullikin[9] showed that

i) $X(z)$ ,$Y(z)$  are analytic  in $|z| >0$ avoiding the essential singularity of the $\exp(-t/z)$ at $z =0-$,

ii) for $0<x<1$, $X^+(x) = X(x)$ , and $X^-(x) = X(x)$ and $Y^+(x) = Y(x)$ , and $Y^-(x) = Y(x)$   where superscript "+" denotes the limiting value of $X(z)$, $Y(z)$  when $z$ approaches  to the cut along $(0,1)$ from upper side and  where superscript "-" denotes the limiting value of $X(z)$ when $z$ approaches  to the cut along $(0,1)$ from lower side.

iii) $X(z)$ , $Y(z)$ are continuous  and  real valued  across the cut along $(0,1)$ with  removable singularity at $z = 1$,

iv) for $0<x<1$, $X^+(-x) = \exp (t/x) Y(x)$ , and $X^-(-x) = \exp(t/x) Y(x)$ , where superscript "+" denotes the limiting value of $X(z)$ when $z$ approaches  to the cut along $(-1,0)$ from upper side and  where superscript "-" denotes the limiting value of $X(z), Y(z)$ when $z$ approaches  to the cut along $(-1,0)$ from lower side.

v) $X(z)$ ,$Y(z)$  are continuous  and real valued across the cut along $(-1,0)$  with removable singularity at  $z= -1$ ,

vi) $X(z)$ , $Y(z)$ have poles at  $z= 1/k$  and  at $z= -1/k$ where $z=+1/k$ and $z=-1 /k$ ,   $0<k \le 1$  are the zeros of dispersion function $T(z)$  in non conservative cases  and also at  infinity when $T(z)$  has double zero at infinity  in conservative cases .

Mullikin [9] also proved that $X(z)$ , $Y(z)$ in equations (17) and(18) will satisfy the linear non homogeneous coupled singular integral equations for  Re $z = u$  ,  $0<u<1$ in the following form :

$$X(u)\ T_c(u) = f_{Xc}(u,t) + u\ P \int_0^1 X(x)\ U(x)\ d\ x\ / (x\ -u) \qquad (22)$$

$$Y(u)\ T_c(u) = f_{Yc}(u,t) + u\ P \int_0^1 Y(x)\ U(x)\ d\ x\ / ( x-u ) \quad . \qquad (23)$$

where $X(u)$ & $Y(u)$ are continuous across the cut along (0,1),

$$f_{Xc}(u,t) = [1 - u \exp(-t/u) \int_0^1 Y(t) U(t) \, d\,t / (t+u)] , \tag{24}$$

$$f_{Yc}(u,t) = \exp(-t/u) [1 - u \int_0^1 X(t) U(t) \, d\,t / (t+u)] , \tag{25}$$

and

$T_c(u)$ will be given by equation (16) and P before the integral indicates the value of the integrals in equations (22) & (23) in Cauchy Principal value sense.

It shall therefore be concluded that real valued $X(u)$ and $Y(u)$ will also satisfy the equations (1) and (2) in addition to equations (22) and (23) with the following constraints at zero $z=1/k$ , $0<k<1$ of $T(z)$ in non conservative cases of $U(x)$ [ i.e $U_0 <1/2$ , $0<k<1$ ]

$$[1 - \exp(-t\,k) \int_0^1 Y(t) U(t) \, d\,t / (k\,t+1)] = \int_0^1 X(x) U(x) / (1-k\,x) \tag{26}$$

$$\exp(-t\,k) [1 - \int_0^1 X(t) U(t) \, d\,t / (k\,t+1)] = \int_0^1 Y(x) U(x) / (1-k\,x) \tag{27}$$

and with following constraints at zeros at infinity of $T(z)$ in conservative cases of $U(x)$ [i.e. $U_0 = \frac{1}{2}$ and $k \rightarrow 0$] :

$$1 = \int_0^1 [X(t) + Y(t)] U(t) \, d\,t \tag{28}$$

$$t \int_0^1 Y(t) U(t) \, d\,t = \int_0^1 [X(t) - Y(t)] \, t \, U(t) \, d\,t \tag{29}$$

Those real valued functions $X(u)$ , $Y(u)$ can be extended to the complex z plane cut along (-1,1) to satisfy the equations (17-18) with the relation

$$Y(z) = \exp(-t/z) \, X(-z) \tag{30}$$

and with zeros of $T(z)$ at $z = 1/k$ and $z = -1/k$ in the complex plane cut along ( -1,1) in non conservative cases and with double zeros at infinity in conservative cases .

### II. Riemann Hilbert Problems:

Our next aim is to use the equations (22) and (23) , to frame a set of Riemann Hilbert problem and to determine a new set linear decoupled integral equations in the complex z plane cut along (-1,1) using the method outlined by Das[ 40] .

Following Muskhelishvili [10] and Das [40] we shall frame two coupled Riemann Hilbert Problems using some functions in the complex z plane cut along (0,1) as follows:

Let

$$P(z,t) = \int_0^1 X(x) \ U(x) / (x-z) \qquad (31)$$

$$Q(z,t) = \int_0^1 Y(x) \ U(x) / (x-z) \qquad . \qquad (32)$$

We shall outline the properties of P(z ,t) as follows :

    i)       $P(z,,t)$ is analytic in the complex plane cut along (0,1) ;

    ii)      $P(z,t) = O(1/z)$ when z $\rightarrow \infty$ ;

    iii)     I P (z, t) I $\leq K_{0p} / $ I z I$^{\alpha}{}_{0p}$ , $0 \leq \alpha_{0p} < 1$ ;

    iv)     I P (z ,t) I $\leq K_{1p} / $ I 1 -z I$^{\alpha}{}_{1p}$ , $0 \leq \alpha_{1p} < 1$;

    v)     U(x) X(x) is continuous in the interval (0,1) and satisfy equivalent Holder conditions and $K_{0p}, K_{1p}, \alpha_{0p}$ and $\alpha_{1p}$ are constants .

We shall also outline the properties of Q(z ,t) as follows :

    i)      $Q(z,t)$ is analytic in the complex plane cut along (0,1) ;

    ii)     $Q(z,t) = O(1/z)$ when z $\rightarrow \infty$ ;

    iii)     I Q (z,t) I $\leq K_{0q} / $ I z I $^{\alpha}{}_{0q}$ , $0 \leq \alpha_{0q} < 1$ ;

    iv)     I Q (z ,t) I $\leq K_{1q} / $ I 1 -z I $^{\alpha}{}_{1q}$ , $0 \leq \alpha_{1q} < 1$;

    v)     U(x) Y(x) is continuous in the interval (0,1) and satisfy equivalent Holder conditions and $K_{0q}, K_{1q}, \alpha_{0q}$ and $\alpha_{1q}$ are constants .

Following the Plemelj's formulae as outlined in Das[40] we can write the equations (22) and (23) we get after some rearrangement,

$$P^+(u,t) = G(u) \ P^-(u,t) + P_x(u,t) \qquad (33)$$

$$Q^+(u,t) = G(u) \ Q^-(u,t) + Q_y(u,t) . \qquad (34)$$

where

$$G(u) = [T_c(u) + i\pi u\ U(u)] / [T_c(u) - i\pi u\ U(u)] \qquad (35)$$

$$P_x(t,u) = \pi i\ U(u)\ f_{xc}(u,t)/ [[T_c(u) - i\ u\ U(u)]] \qquad (36)$$

$$Q_y(t,u) = \pi i\ U(u)\ f_{yc}(u\ .t)/ [[T_c(u) - i\ u\ U(u)]] \qquad (37)$$

$f_{xc}(u, t)$ , $f_{yc}(u, t)$ are given by equations (24) and (25) .
$G(u)$ in equation (35) is a known complex function .

$$T^+(u) = T_c(u) + i\pi\ u\ U(u) \qquad (38)$$

$$T^-(u) = T_c(u) - i\pi\ u\ U(u) \qquad (39)$$

$$G(u) = T^+(u) / T^-(u) \qquad (40)$$

where $T_c(u)$ is given by equation (16).

From equations (33) and (34) we shall determine the unknown function $P(z,t)$ , $Q(z,t)$ for being analytic in the complex plane cut along (0,1) having properties outlined above and $G(u)$ and $U(u)$ are known functions. These are known as non homogeneous Riemann Hilbert problem.

### III. Solutions of Riemann Hilbert Problems :

We shall first determine the solution of homogeneous Riemann Hilbert problems obtained from equations (33) and (34) using the method in Das [40]

It is that of finding a new analytic function $N(z)$ for which

$$N^+(u) = G(u)\ N^-(u) , \quad 0 < u < 1 \qquad (41)$$

where $N(z)$ is to satisfy the following properties :

i)      $N(z)$ is analytic in the complex plane cut along (0,1) ;

ii)     $N(z)$ is not zero for all z in complex plane cut along (0,1) ;

iii)    $O(1/z)$ when $z \to \infty$ ;

iv)     $1\ N(z)\ 1 \le K_2/\ 1\ z\ 1^{\alpha_2} , \quad 0 \le \alpha_2 < 1$ ;

v)      $1\ N(z)\ 1 \le K_3/\ 1\ 1\ -z\ 1^{\alpha_3} , 0 \le \alpha_3 < 1$ , where $K_2$, $K_3$ , $\alpha_2$ and $\alpha_3$ are constants .

We now assume that for $0 < u < 1$, $U(u)$ is real , positive , single valued and satisfy Holder condition in (0,1) . We also assume that $T_c(u)$ is not equal to zero both in the conservative case $U_0 = \frac{1}{2}$ and in non conservative cases $U_0 < \frac{1}{2}$ .

As $T_c(u)$ is dependent on $U(u)$ it can be proved that $T_c(u)$ is real ,one valued and satisfy Holder conditions in (0,1) and modulus of ( $T_c(u) + i \pi u U(u)$ ) , ( $T_0(u) - i \pi u U(u)$ ) are not equal to zero, $T_c(0) = 1$ and $T_c(1) \rightarrow -\infty$ as $u \rightarrow 1$ from within the interval (0,1) .

Taking logarithm to equation (41) and using equations (38) , (39) and (40) we find that

$$\log N^+(u) - \log N^-(u) = \log G(u) = 2 i \theta (u) \qquad (42)$$

where

$$\tan [ \theta(u) ] = \pi u U (u) / T_c(u) \quad ; -\pi/2 < \theta (u) < \pi/2 \qquad (43)$$

$\theta(u)$ is assumed single valued .

Here we can take

$\theta(0) = 0,$

$\theta(1) = r \pi$ , $r = 1,2,3$ etc $\qquad (44)$

$r = \frac{1}{2}$ the zeros of $T(z)$ in the complex plane cut along (-1,1)

$= 1$ in this case .

We shall now evaluate $N(z)$. Using Plemelj's formulae to equation (42) we get

$$\log N(z) = (2\pi i)^{-1} \int_0^1 \log G(x) d x / (x - z) + P_n (z) . \qquad (45)$$

where $P_n(z)$ is an arbitrary polynomial in z of degree n. in the complex plane cut along (0,1) and it is continuous in the complex plane across a cut along (0,1) because it is analytic there. Thus equation (45) does also satisfy the Plemelj's formulae.

We set

$$L(z) = (2\pi i)^{-1} \int_0^1 \log G(u) d u / ( u - z ) \qquad (46)$$

Equation (45) with equation (46) then yields

$$N(z) \ = \ N_0 \, (z) \ \ exp \, ( \ L(z) \ )  \hspace{3cm} (47)$$

where $N_0(z) = exp \, ( \ P_n(z) \ )$ is analytic for all z in the complex plane and is to be determined such that N(z) given by equation (47) satisfy all properties outlined above and the use of the Plemelj's formulae is not invalidated .

We can determine the value of $N_0(z)$ at the end points u=0 and u=1 using the method outlined by Das [40] as

$$N_0(z) \ = (1\text{-}z)^{-1}  \hspace{3cm} (48) \, .$$

Hence equation (47) using equation (48) , yields

$$N(z) \ = ( \ 1 -z \ )^{-1} \ \ exp(L(z))  \hspace{2cm} (49).$$

This is the solution of homogeneous Hilbert problem mentioned in equation (41) in the complex z plane cut along (0,1) .

This can be written using equations (42) , (43), (46) , (47) in explicit form when z is in complex in z plane cut along(0,1))

$$N(z) \ = \ (1\text{-}z)^{-1} \ \ exp \, ( \ \pi^{-1} \int_0^1 \ \theta(u) \, d \, u \, / \, (u - z) \ )  \hspace{1cm} (50)$$

The solution of the homogeneous Hilbert problem arising from equations (33) and (34) using equation (41) is called the complementary solution of the non homogeneous Riemann Hilbert Problem as :

$$P_c(z,t) \ = \ A \ N(z)  \hspace{3cm} (51)$$

$$Q_c(z,t) = \ B \ N(z)  \hspace{3cm} (52)$$

where A and B are arbitrary constants to be determined using the zeros of dispersion function T(z) and /or conditions at z→ 0 and /or conditions at infinity .

Equations (33) & (34) with equation (41) give

$$P^+(u,t) \, / \, N^+(u) - P^-(u,t) \, / \, N^-(u) = 2\pi i \ U(u) \ f_{xc}\,(u,t) \, / \, N^+(u) \ T^-(u) \, ,  \hspace{0.5cm} (53)$$

$$Q^+(u,t) \, / \, N^+(u) - Q^-(u,t) \, / \, N^-(u) = 2\pi i \ U(u) \ f_{yc}\,(u,t) \, / \, N^+(u) \ T^-(u) \, ,  \hspace{0.5cm} (54)$$

when u lie in the interval (0,1).

From equations (53) and (54) we get for z in complex plane cut along (0,1), the particular solutions as

$$P_p(z,t) \ = \ N(z) \int_0^1 \ U(u) f_{xc}(u,t) \, d \, u \, / \, (N^+(u) \ T^-(u) \, ( \, u - z) \ )$$

$$+ \quad P_{mx}(z) \ N(z) \qquad\qquad (55)$$

where $P_{mx}(z)$ is an arbitrary polynomial in z of degree m and is continuous in the complex plane across a cut along (0,1 ) and

$$Q_p(z,t) \ = \ N(z) \int_0^1 \ U(u) f_{yc}(u,t) \, d \, u \, / \, (N^+(u) \ T^-(u) \ ( \, u - z \, ) \ )$$

$$+ \quad P_{py}(z) \ N(z) \qquad\qquad (56)$$

where $P_{py}(z)$ is an arbitrary polynomial in z of degree p and is continuous in the complex plane across a cut along (0,1 )

Thus equations (55) & (56) do satisfy the Plemelj's formulae . The arbitrariness of $P_{mx}(z)$ and $P_{py}(z)$ are removed usually by examining the behaviour of P(z,t) ,Q(z,t) and N(z) at infinity and /or the end points of the cut along (0,1) .

Since N(z) is equivalent to $\exp(L(z)$ ) , the term appearing on the RHS of equations (55) and (56) are polynomials of degree m and p respectively . However, if we use the properties of N(z) and P(z, t) and Q(z, t) when z→ ∞,

we must have

$$P_{mx}(z) = 0 \qquad\qquad\qquad (57)$$

$$P_{py} (z) = 0 \qquad\qquad\qquad (58)$$

Hence we get the particular solution $P_p(z,t)$, $Q_p(z,t)$ of equations (33) & (34) in complex z plane cut along (0,1)) for P(z, t) and Q(z,t) as

$$P_p(z, t) \ = \ N(z) \int_0^1 \ U(u) f_{xc}(u, t) \, d \, u \, / \, (N^+(u) \ T^-(u) \ ( \, u - z \, ) \ ) \qquad (59).$$

$$Q_p(z,t) \ = \ N(z) \int_0^1 \ U(u) f_{yc}(u,t) \, d \, u \, / \, (N^+(u) \ T^-(u) \ ( \, u - z \, ) \ ) \qquad (60)$$

.P (z, t) and Q(z ,t) in equations (31) and (32) will then be the sum of complementary solution of homogeneous Hilbert problem obtained from equations (51) and (52) and the particular solution obtained from the equations (59)and (60).

Hence the equations in (31) and (32) become

$$P(z,t) = \int_0^1 X(x) \, U(x) / (x - z)$$

$$= A \, N(z) + N(z) \int_0^1 U(u) f_{xc}(u,t) \, du / (N^+(u) \, T^-(u) \, (u-z))$$

$$\tag{61}$$

$$Q(z,t) = \int_0^1 Y(x) \, U(x) / (x - z) \quad .$$

$$= B \, N(z) + N(z) \int_0^1 U(u) \, f_{yc}(u,t) \, du / (N^+(u) \, T^-(u) \, (u-z))$$

$$\tag{62}$$

where $N(z)$ and $T^-(u)$ are given by equations (50) and (39) and $N^+(u)$ will be determined from $N(z)$ as

$$N^+(u) = (1 - u)^{-1} \exp(L^+(u)), \tag{63}$$

$$L^+(u) = (\pi)^{-1} P \int_0^1 \theta(t) \, dt / (t-u) + i \, \theta(u) \quad , \tag{64}$$

P before the integral (64) indicates the value of the integral in Cauchy Principal value sense.

The use of equations (61) & (62) to equations (17) & (18) will give linear coupled integral equations for $X(z)$ and $Y(z)$ in the complex cut along $(-1,1)$ as

$$X(z) \, T(z) = f_{xc}(z,t) + z \, N(z) \, [A +$$
$$\int_0^1 U(u) f_{xc}(u,t) \, du / (N^+(u) \, T^-(u) \, (u-z))] \tag{65}$$

$$Y(z) \, T(z) = f_{yc}(z,t) + z \, N(z) [B +$$
$$\int_0^1 U(u) f_{yc}(u,t) \, du / (N^+(u) \, T^-(u) \, (u-z))] \tag{66}$$

where $f_{xc}(z,t)$ & $f_{yc}(z,t)$ are given by equations (19) and (20).

## IV. Decoupled X- and Y- functions :

We shall now analytically continue the result in equations (61) and (62) in the complex plane cut along $(-1,0)$ by replacing z by $-z$, as

$$\int_0^1 X(x) \ U(x) \ / \ (x + z)$$

$$= A \ N \ (-z) + \ N \ (-z) \int_0^1 \ U \ (u) f_{xc}(u, t) \ d \ u \ / \ (N^+(u) \ T^-(u) \ ( \ u + z) \ )$$

$$(67)$$

$$\int_0^1 Y(x) \ U(x) \ / \ (x + z) \quad .$$

$$= B \ N \ (-z) + N \ (-z) \int_0^1 \ U \ (u) \ f_{yc}(u \ ,t) \ d \ u \ / \ (N^+(u) \ T^-(u) \ ( \ u + z) \ )$$

$$(68)$$

Let

$$\Phi_1(-z,t) \ = \ \int_0^1 \ U \ (u) f_{xc}(u, t) \ d \ u \ / \ (N^+(u) \ T^-(u) \ ( \ u + z) \ ) \qquad (69)$$

$$\Phi_2(-z,t) \ = \ \int_0^1 \ U \ (u) \ f_{yc}(u \ ,t) \ d \ u \ / \ (N^+(u) \ T^-(u) \ ( \ u + z) \ ) \qquad (70)$$

Those $\Phi_1(-z,t)$ , $\Phi_2(-z,t)$ are analytic in the complex z plane cut along $(-1,0)$. On substitution of $f_{xc}(u,t)$ and $f_{yc}(u,t)$ from equations(19) and (20) in the equations (69) & (70) and using

$$1 \ / \ [(y +x) \ (x +z) \ ] = [ \ y/(x +y) \ - z/ \ (x +z) \quad ] \ / \ [x \ (y-z) \ ] \qquad (71)$$

we can find after some manipulation that

$$\Phi_1(-z \ ,t) \ = N_1(-z) + N_2(-z \ ,t \ ) \int_0^1 \ Y(y) \ U(y) \ d \ y \ / \ (y \ - z)$$

$$- \int_0^1 \ Y(y) \ N_2(-y \ ,t \ ) \ U(y) \ d \ y \ / \ (y \ - z) \quad , \qquad (72)$$

$$\Phi_2(-z \ ,t) \ = [ \ N_2(-z, t) \ / \ z \ ] + N_2(-z \ ,t \ ) \int_0^1 \ X(y) \ U(y) \ d \ y \ / \ (y \ - z)$$

$$- \int_0^1 \ X \ (y) \ N_2 \ (-y \ ,t \ ) \ U(y) \ / \ (y \ - z) \quad , \qquad (73)$$

where $N_1(-z)$ and $N_2(-z \ ,t)$ are defined as

$$N_1(-z) = \int_0^1 U(u)d\,u\,/\,(N^+(u)\,T^-(u)\,(u+z)\,) \tag{74}$$

$$N_2(-z,t) = z\int_0^1 U(u)\exp(-t/u)\,d\,u\,/\,(N^+(u)\,T^-(u)\,(u+z)\,)\,. \tag{75}$$

Equations (67) & (68) with equations (69-73) becomes

$$\int_0^1 X(x)\ U(x)\,/\,(x+z)$$

$$= A\,N(-z) + N(-z)\,\Phi_1(-z,t)$$

$$= A\,N(-z) + N(-z)[\,N_1(-z) + N_2(-z,t)\int_0^1 Y(y)\ U(y)\,d\,y\,/\,(y-z)$$

$$-\int_0^1 Y(y)\,N_2(-y,t)\,U(y)\ d\,y\,/\,(y-z)\,]\,, \tag{76}$$

$$\int_0^1 Y(x)\ U(x)\,/\,(x+z)\quad.$$

$$= B\,N(-z) + N(-z)\,\Phi_2(-z,t)$$

$$= B\,N(-z) + N(-z)[\,[\,N_2(-z,t)\,/\,z\,]$$

$$+ N_2(-z,t)\int_0^1 X(y)\ U(y)\ d\,y\,/\,(y-z)$$

$$-\int_0^1 X(y)\,N_2(-y,t)\,U(y)\,/\,(y-z)\,]\,, \tag{77}$$

Equations (76) and (77) give respectively integrals of X(z) in terms of Y(z) only and integrals of Y(z) in terms of X(z) only and those are thereby decoupled.

We shall now use equations (76) and (77) in equations (19) and (20) to get $f_{xc}(z,t)$ and $f_{yc}(z,t)$ in terms of integrals of X(z) and Y(z) respectively as follows:

$$f_{xc}(z,t) = 1 - z\exp(-t/z)\ N(-z)\,[\,[\,B + N_2(-z,t)\,/\,z\,]$$

$$+ N_2(-z,t)\int_0^1 X(y)\ U(y)\ d\,y\,/\,(y-z)$$

$$-\int_0^1 X(y)\,N_2(-y,t)\,U(y)\,/\,(y-z)\,]\,, \tag{78}$$

$f_{yc}(z, t) = \exp(-t/z) - z \exp(-t/z) \ N(-z) \ [ \ [ \ A \ + N_1(-z) \ ]$

$+ \ N_2(-z, t) \int_0^1 \ Y(y) \ U(y) \ d \ y \ / \ (y \ - z)$

$- \int_0^1 \ Y(y) \ N_2(-y, t) \ U(y) \ d \ y \ / \ (y \ - z) \ ]$ .    (79)

We shall now use the expressions for $f_{xc}(z, t)$ and $f_{yc}(z, t)$ from equations (78) & (79) to equations (17) & (18) to get linear non homogeneous decoupled new integral equations (LNDIE) in the complex z plane cut along (-1,1) for $X(z)$ and $Y(z)$ as follows

$X(z) \ T(z) = [1 - z \exp(-t/z) \ N(-z) \ [ \ B + N_2(-z, t) \ / \ z \ ] ]$

$+ \ z \int_0^1 \ X(x) \ U( \ x, -z, \ t) \ d \ x \ / \ (x - z)$ ,    (80)

$Y(z) \ T(z) = [ \ \exp(-t/z) - z \exp(-t/z) \ N(-z) \ [ A + N_1(-z) ] \ ] ]$

$+ \ z \int_0^1 \ Y \ (x) \ U(x, -z, \ t) \ d \ x \ / \ (x - z)$ .    (81)

where

$\quad U \ (x, -z, t) = U(x) \ [ \ 1 - N(-z) \ N_2(-z, t) \ \exp(-t/z)$

$\qquad + \ N(-z) \ N_2(-x, t) \ \exp(-t/z) \ ]$    (82)

$\quad U(x, -x, \ t) \ = \ U \ (x)$    (83)

where constraints are

$[1 - \ k^{-1} \exp(-t \ k) \ N(-1/k) \ [ \ B + N_2(-1/k, t) \ k] \ ]$

$= \ \int_0^1 \ X(x) \ U( \ x, -1/k, \ t) \ d \ x \ / \ (1 - k \ x)$ ,    (84)    $[$

$\exp(-t \ k) - \ k^{-1} \ \exp(-t \ k) \ N(-1/k) \ [ \ A + N_1(-1/k) \ ] \ ]$

$= \ \int_0^1 \ Y \ (x) \ U(x, -1/k, t) \ d \ x \ / \ (1 - k \ x)$ .    (85)

## V. New functions $N_1(z)$ and $N_2(z, t)$ in terms of $N(z)$:

We shall now represent, following the procedure outlined in Das[40], $N_1(z)$ in terms of $N(z)$ in the complex z plane cut along (0,1) by using Cauchy's integral theorems. We get from equation (74)

$N_1(z) = \int_0^1 \ U(u) d \ u \ / (N^+(u) T^-(u) \ (u - z)$    (86)

where $N_1(z)$ is analytic in the complex z plane cut along (0, 1).

We consider contour integrals as follows:

$$F(z) = (2\pi i)^{-1} \int_{C_1} F(w) \, dw \, / \, (w-z) ) \qquad (87)$$

where $C_1 = L_1 \cup L_2$ . Here contour $L_1$ is a sufficiently large in the complex z plane to contain the cut along $(0,1)$ within and $L_2$ is a circle with center at the origin and of very large radius R and $L_1$ lies interior to this $L_2$ and z lies outside $L_1$ but inside $L_2$ . Both $L_1$ and $L_2$ are taken in counter clockwise sense. The function $N(z)$ and $1/N(z)$ are analytic and nonzero in the annulus $C_1$ . We shall now apply the Cauchy Integral theorem on the function

$F(z) = 1/ [ z N(z) ]$ around $C_1$ to obtain

$$1/ [z N(z)] = (2\pi i)^{-1} \int_{L_2} dw \, / \, ( N(w) \, w \, (w-z) )$$

$$- (2\pi i)^{-1} \int_{L_1} dw \, / \, ( N(w) \, w \, (w-z) ) , \qquad (88)$$

Using equation (49) we can write

$$1/N(w) = ( 1 - w ) \exp ( - L (w) ) \qquad (89)$$

and when $w = u$ , u real and $0 < u < 1$

we can write from equation.(89)

$$1/N^{+}(u) = ( 1 - u ) \exp ( - L^{+} (u) ) \qquad (90)$$

$$1/N^{-}(u) = ( 1 - u ) \exp ( - L^{-} (u) ) \qquad (91)$$

where the superscript '+' denotes the value of functions for taking $w = u$ from above the cut along $(0,1)$ and the superscript '-' denotes the value of functions for taking $w = u$ from below the cut along $(0,1)$ of the respective functions in the complex z plane and

$$L^{+}(u) = ( \pi^{-1} P \int_0^1 \theta(t) \, dt \, / \, ( t - u ) + i \, \theta (u) ) \qquad (92)$$

$$L^{-}(u) = ( \pi^{-1} P \int_0^1 \theta(t) \, dt \, / \, ( t - u ) - i \, \theta (u) ) \qquad (93)$$

Using equations (92) , (93) , ( 38) , (39) and (40) in equations (90) and (91

we get

$$1 / N^{+}(u) - 1 / N^{-}(u) = - 2\pi i \, u \, U (u) \, / \, T^{-}(u) \, N^{+}(u) \qquad (94)$$

To make the contour $L_1$ to be well defined we shall shrink the contour $L_1$ to

i)     a  circle  $C_0$  around the origin of the complex plane  in  counter clockwise sense such that  w= r exp (i α), 0≤ α ≤ 2π ;

ii)     a line CD below the cut along (0,1)   from  r to 1-r where w = u, real  , u varies from 0 to 1;

iii)     a circle $C_2$  counter clockwise sense around  w=1  such that  w = 1  + r exp(i β) , -π ≤ β≤ π  and

iv)     a line BA above the cut  along (0,1) from 1-r to r  on which   w = u , real ,  u varies from 1 to 0 .

The value of the integral , on the circle $C_0$ ,  in the limit r➔0   becomes

$$(2\pi i)^{-1} \int_{C_0}  d\, w\, / \,(\,N(w)\,w\,(w-z)\,\,) \,=\, -\,(\,z\,N(0)\,)^{-1} \quad (95)\,.$$

The value of the integral , on the circle $C_2$, in the limit r➔0 becomes

$$(2\pi i)^{-1} \int_{C_2}  d\, w\, / \,(\,N(w)\,w\,(w-z)\,\,) \,\, = \,\, 0 \qquad\qquad (96)$$

The value of the integral ,on the line CD , in the limit r➔0 becomes

$$(2\pi i)^{-1} \int_{CD}  d\, w\, / \,(\,N(w)\,w\,(w-z)) = (2\pi i)^{-1} \int_0^1  d\, u\, / \,(\,N^-(u)\,u\,(u-z)\,\,), \,(97)$$

The value of the integral , on the line BA ,  in the limit r➔0   becomes

$$(2\pi i)^{-1} \int_{BA}  d\, w\, / \,(\,N(w)\,w\,(w-z)) = (2\pi i)^{-1} \int_1^0  d\, u\, / \,(\,N^+(u)\,u\,(u-z)\,\,) \,, \,(98)$$

Hence the integral on the contour $L_1$ in equation (88), using equations (94-98) becomes

$$(2\pi i)^{-1} \int_{L_1}  d\, w\, / (N(w)\,w\,(w-z)\,)$$

$$= \int_0^1  U(u)\, d\, u\, / (\,N^+(u) T^-(u)\,(u-z)\,)$$

$$-\,1\,/\,(\,z\,N(0)\,) \qquad\qquad (99)$$

The integral , on the contour $L_2$ , when  R➔ ∞   becomes

$$(2\pi i)^{-1} \int_{L2} d\,w / (N(w)\,w\,(w-z)\,) = -1 \qquad (100)$$

as $w\,N(w) = -1$ when $w \to \infty$ $\qquad (101)$

Hence equation (88) with equations (99), (100) and (101) gives

$$N_1(z) = (\,z\,N(0)\,)^{-1} - 1 - (z\,N(z))^{-1} \qquad (102)$$

$$N_1(-z) = -(\,z\,N(0)\,)^{-1} - 1 + (z\,N(-z))^{-1} \qquad (103)$$

We shall now determine $N_2(z,t)$ in terms of $N(z)$.
Let us set

$$V_2(z\,,t) = \int_0^1 U(u)\exp(-t/u)\,d\,u / (N^+(u)\,T^-(u)\,(\,u-z\,)\,). \qquad (104)$$

where $V_2(z,t)$ is analytic in the complex $z$ plane cut along $(0,1)$.

We consider a new contour integral as follows :

$$F(z) = (2\pi i)^{-1} \int_{C1} F(w)\,d\,w / (w-z)\,) \qquad . \qquad (105)$$

where $C_1 = L_1 \cup L_2$ . Here contour $L_1$ is a sufficiently large in the complex $w$ plane to contain the cut along $(0,1)$ within and $L_2$ is a semicircle with center at the origin and of very large radius R in Re $z > 0$ to avoid the essential singularity at $z \to 0-$ with diameter on the imaginary axis and $L_1$ lies interior to this $L_2$ and z lies outside $L_1$ but inside $L_2$ . Both $L_1$ and $L_2$ are taken in counter clockwise sense. The functions $N(w)$ and $1/N(w)$ are analytic and nonzero in the annulus $C_1$ . We shall now apply the Cauchy Integral theorem on the function

$$F(w) = \exp(-t/w\,) / [\,w\,N(w)\,] \qquad (106)$$

around $C_1$ to obtain

$$\exp(-t/z) / [z\,N(z)] = (2\pi i)^{-1} \int_{L2} \exp(-t/w)\,d\,w / (\,N(w)\,w\,(w-z)\,)$$

$$- (2\pi i)^{-1} \int_{L_4} \exp(-t/w)\,d\,w / (\,N(w)\,w\,(w-z)\,) \qquad . \qquad (107)$$

To make the contour $L_1$ to be well defined we shall shrink the contour $L_1$ to

i)       a line CD below the cut along (0,1)   from  0+ to 1-r where w = u, real , u varies from 0 to 1;

ii)      a circle $C_{11}$ counter clockwise sense around  w=1  such that  w = 1 + r exp(i β ) , -π ≤ β≤ π  and

iii)     a line BA above the cut  along (0,1) from 1-r to 0+ on which   w = u , real ,  u varies from 1 to 0 .

The value of the integral , on the circle $C_{11}$ , in the limit r→0   becomes

$$(2\pi i)^{-1} \int_{c11} \exp(-t/w)\, dw / (\ N(w)\, w\, (w-z)\ ) = 0 \quad .(108).$$

The value of the integral ,on the line CD , in the limit r→0 becomes

$$(2\pi i)^{-1} \int_{CD} \exp(-t/w)\, dw / (\ N(w)\, w\, (w-z)\ )$$

$$= (2\pi i)^{-1} \int_0^1 \exp(-t/u)\, du / (\ N^+(u)\, u\, (u-z)\ ) \qquad (109).$$

The value of the integral ,on the line BA  , in the limit r→0 becomes

$$(2\pi i)^{-1} \int_{BA} \exp(-t/w)\, dw / (\ N(w)\, w\, (w-z)\ )$$

$$= - (2\pi i)^{-1} \int_0^1 \exp(-t/u)\, du / (\ N^-(u)\, u\, (u-z)\ ) \qquad (110).$$

.

To make the contour $L_2$ to be well defined we shall shrink the contour    $L_2$ following Dasgupta [27] to

i)       a line CP along  the imaginary axis on which  w = -i y  where y varies from r to R ;

ii)      a semicircle $D_{11}$ counter clockwise sense around  w=0+  such that w =  R exp(i α ) ,  -π/2 ≤ α ≤ π/2  in  Re w >0  and

iii)     a line QB along the imaginary axis on which  w =  i y where   y varies from  R to r.

The value of the integral ,on the line CP , in the limit r→0 and R →∞ becomes

$$(2\pi i)^{-1} \int_{CP} F(w)\, dw / (w-z) = (2\pi i)^{-1} \int_0^{-i\infty} F(w)\, dw / (\ w -z\ ) , (111)$$

The value of the integral ,on the line QB , in the limit r→0 and R →∞ becomes

$$(2\pi i)^{-1} \int_{QB} F(w) \, dw \, / \, (w-z) = (2\pi i)^{-1} \int_{i\infty}^{0} F(w) \, dw \, / \, (w-z) \quad ,(112)$$

For large w , using equation (106) we have ,

$$F(w) = -1 + O(1/w) \quad . \tag{113}$$

Hence the value of the integral on the semi-circle $D_{11}$ in the limit of R→∞ becomes

$$(2\pi i)^{-1} \int_{D11} F(w) \, dw \, / \, (w-z) = -1/2 \tag{114}$$

Hence equation (107), using equations (108-114) gives

$$\int_{0}^{1} U(u) \exp(-t/u) \, du \, / \, (N^{+}(u) \, T^{-}(u) \, (u - z)) = - \exp(-t/z) \, / \, [z \, N(z)] \, -1/2$$

$$+ (2\pi i)^{-1} \int_{i\infty}^{0} F(w) \, dw \, / \, (w-z) + (2\pi i)^{-1} \int_{0}^{-i\infty} F(w) \, dw \, / \, (w-z)$$

$$= - \exp(-t/z) \, / \, [z \, N(z)] \, -1/2 \, - (2\pi i)^{-1} \int_{-i\infty}^{+i\infty} F(w) \, dw \, / \, (w-z)$$

$$\tag{115}$$

We can write

$$V_{20}(z,t) = (2\pi i)^{-1} \int_{-i\infty}^{+i\infty} F(w) \, dw \, / \, (w-z)$$

$$= (2\pi i)^{-1} \int_{-i\infty}^{+i\infty} \exp(-t/w) \, dw \, / \, (w \, N(w) \, (w-z)) \tag{116}$$

where $V_0(z, t)$ is analytic in Re z > 0 and 0 (1/z) when z →∞

Hence equation(104) with equations (115-116) gives

$$V_2(z,t) = - \exp(-t/z) \, / \, [z \, N(z)] - 1/2 - V_{20}(z,t) \quad , \tag{117}$$

Equation (75) with equations (100) ,(117) gives $N_2(-z,t)$ and $N_2(z,t)$ in terms of N(z) as

$N_2(-z, t) = z\, V_2(-z, t)$

$$= \exp(t/z) / N(-z) \quad - z/2 \; - z\, V_{20}(-z, t) \qquad (118)$$

$N_2(z, t) = - z\, V_2(z, t)$

$$= \exp(-t/z) / N(z) \quad + z/2 \; + z\, V_{20}(z, t) \qquad (119)$$

## VI.  Riemann Hilbert problems for decoupled  X- and Y- functions:

If we use the expression of $N_2(-z, t)$  from equation (118) and $N_1(-z)$ from equation (103)   in the non homogeneous part of the decoupled equations (80) and (81) , we then get

$f_{Xd}(z, t) = 1 - z \exp(-t/z)\, N(-z) \; [\; B + N_2(-z, t) / z \;]$

$$= z \exp(-t/z)\, N(-z) \; [\; 1/2 \; - B \; + V_{20}(-z, t) \;] \;, \qquad (120)$$

$f_{Yd}(z, t) = \exp(-t/z) - z \exp(-t/z)\, N(-z) \; [\; A + N_1(-z) \;]$

$$= z \exp(-t/z)\, N(-z) \; [\; 1 - A + (z\, N(0))^{-1} \;] \;, \qquad (121)$$

Equation (80) & (81) with equations (120) & (121) give the simplest form of the linear decoupled integral equations  for X(z) and Y(z)  as follows :

$$X(z)\, T(z) = f_{Xd}(z, t)$$

$$+ z \int_0^1 X(x)\, U(x, -z, t)\; dx\; / (x - z) \qquad , \qquad (122)$$

$$Y(z)\, T(z) = f_{Yd}(z, t)$$

$$+ z \int_0^1 Y(x)\, U(x, -z, t)\, dx\; / (x - z) \quad . \qquad (123)$$

with constraints

$$k^{-1} \exp(-t\, k)\, N(-1/k) \; [\; 1/2 \; - B \; + V_{20}(-1/k, t) \;]$$
$$= \int_0^1 X(x)\, U(x, -1/k, t)\; dx\; / (1 - k\, x) \quad , \qquad (124)$$

$$k^{-1} \exp(-t\, k)\, N(-1/k) \; [\; 1 - A + k\, (N(0))^{-1} \;]$$
$$= \int_0^1 Y(x)\, U(x, -1/k, t)\; dx\; / (1 - k\, x) \quad , \qquad (125)$$

The equations (122) and (123) can be written in one compact form as

$$Q(z)\ T(z)\ =\ f\ (z,t)\ +\ z\int_0^1\ Q(x)\ U(\ x,\ -z\ ,\ t)\ d\ x\ /\ (x-z)\quad (126)$$

where z lies in the cut plane along (-1,1) and

Q (z) = either X (z) or Y (z) as the case when z is outside the cut along (-1,1) and

Q (x) = either X (x) or Y(x) as the case may be when z = x on the cut along (0,1)and f (z, t)= $f_{Xd}$ (z ,t) and $f_{yd}$ (z ,t) as the case may be for X(x) and Y(z) respectively as they are decoupled .

Following Mullikin [9] X(x) and Y(x) are real valued for $0 < x < 1$ and continuous across the cut along (0.1) and X(x) , Y(x) can be extended meromorphically in the complex domain Re (z) >0 to X(z) and Y(z) respectively. Accordingly , Q(x) may be extended meromorphically to Q(z) in the complex domain Re (z) >0 .
The equation (126) is a linear decoupled integral equations This can be changed to linear decoupled singular integral equations to the form using Plemelj's formulae.

$$Q(x)\ T_c(x)\ =\ f\ (x,t) + x\ P\int_0^1\ Q(y)\ U(\ y,\ -x\ ,\ t)\ d\ y\ /\ (y-x)\quad .\ (127)$$

where $0 \le x \le 1$ .P indicates Cauchy principal value sense.

We can set

$$R(z\ ,\ t)\ =\ \int_0^1\ Q(y)\ U(\ y,\ -z\ ,\ t)\ d\ y\ /\ (y-z)\ ,\qquad (128)$$

where z is complex .
Applying Plemelj's formulae in equation (128) when z approaches to y on the cut along (0,1) from above and below and substituting those in equation(127) , after some rearrangements, to get

$$R^+(y\ ,t) = R^-(y\ ,\ t)\ T^+(y)\ /\ T^-(y)$$

$$+\ 2\pi i\ U(y,-y\ ,t)\ f\ (y\ ,t)\ /\ T^-(y)\ ,\ 0 < y < 1,\qquad (129)$$

Equation (129) is the new Riemann Hilbert problem . Here N(z) is a function already determined in equation(50) and $N^+(y)$ and $N^-(y)$ are the values of N(z) on the cut along (0,1) when z approaches to y from above and below the cut respectively .We use   equations ( 38-41)  to equation (129) to take the new form  as

$$R^+(y, t) / N^+(y)  =  R^-(y, t) / N^-(y)$$

$$+ 2\pi i \, U(y, -y, t) \, f(y, t) / N^+(y) T^-(y) , \; 0 < y < 1, \qquad (130)$$

We shall outline the properties of R(z ,t) as follows  :

vi)      R(z, , t ) is analytic in the complex plane cut along (0,1) ;

vii)      R(z, t ) = O (1/z)  when z $\rightarrow \infty$ ;

viii)      I R (z, t) I  $\leq$  $K_{0r}$ /  I z I$^{\alpha_{0r}}$ ,  $0 \leq \alpha_{0r} < 1$  ;

ix)      I R (z ,t)  I $\leq$ $K_{1r}$ /  I 1 -z 1 $^{\alpha_{1r}}$ , $0 \leq \alpha_{1r} < 1$;

x)      U(x,-x ,t) Q(x)  is  continuous in the interval (0,1) and satisfy equivalent Holder  conditions    and $K_{0r}$, $K_{1r}$ , $\alpha_{0r}$ and $\alpha_{1r}$ are constants .

We shall determine  the solution of the non  homogeneous Hilbert problem arising from equations  (130) using the same procedure in article (3).

## VII. Solution  of  Riemann Hilbert  problems for decoupled   X- and Y-functions:

The homogeneous Hilbert problems arising from equations (130)  is of finding solutions of

$$R^+ ( \, y, t) N^+(y) =  N^- (y)  R^- (y ,t) \qquad (131)$$

Using the  procedure outlined in Das[40] and  in article (V ) herein , the solution of Homogeneous Hilbert Problem  is

$$R_c(z,t)  =  A_R  N(z) \qquad (132)$$

where  $A_R$ is arbitrary constant  to be determined using the zeros of T(z) and /or conditions at  z$\rightarrow$ 0 and/or conditions at infinity .

We shall now determine particular solutions of non homogeneous Hilbert problems (130) using the procedure in Article (V) :as

$$R_p(z ,t)  =  N(z) \int_0^1  U(y) f(y, t) \, d \, y / (N^+(y) \, T^-(y) \, ( \, y - z) \, )$$

$$+ \quad P_{mR}(z) \; N(z) \hspace{4cm} (133)$$

where $P_{mR}(z)$ is an arbitrary polynomial in z of degree m and is continuous in the complex plane across a cut along (0,1) .

Thus equation (133) satisfies the Plemelj's formulae . The arbitrariness of $P_{mR}(z)$ is removed usually by examining the behaviour of R(z,t) and N(z) at infinity and /or the end points of the cut along (0,1) .

Since N(z) is equivalent to exp(L(z)) , the term appearing on the RHS of equation (133) is a polynomial of degree m . However, if we use the properties of N(z) and R(z, t) when z→ ∞, we must have

$$P_{mR}(z) = 0 \hspace{5cm} (134)$$

Hence we get the particular solution $R_p(z ,t)$ of equations (130) in complex z plane cut along (0,1)) for R(z, t) as

$$R_p(z, t) \;\; = \;\; N(z) \int_0^1 \; U(y) f(y, t) \, d \, y \, / \, (N^+(y) \; T^-(y) \, ( \, y - z) \; ) \; . \hspace{0.5cm} (135).$$

Hence the equations (128) with equations (132) and (135) will take a new form the complex z plane cut along (0,1) as follows:

$$R(z ,t \; ) \; = \; \int_0^1 \; Q(y) \; U(y,-z ,t) \, / \, (y \, - z)$$

$$= \; A_R \, N \, (z)$$

$$+ \, N \, (z) \int_0^1 \; U \, (y,-z ,t) f(y, t) \, d \, y \, / \, (N^+(y) \; T^-(y) \, ( \, y - z) \; ) \; . \; (136)$$

where N(z) and $T^-(y)$ are given by equations ( 50) and ( 39 ) respectively and $N^+(u)$ will be determined from N(z) as

$$N^+ (y) \; = ( \, 1 - y) ^{-1} \exp( \, L^+(y)), \hspace{3cm} (137)$$

$$L^+(y) \; = (\pi)^{-1} P \int_0^1 \; \theta \, (t) \, d \, t \, / \, ( \, t - y) \; + i \; \theta \, (y) \; , \hspace{2cm} (138)$$

P before the integral (138) indicates the value of the integral in Cauchy principal value sense.

The use of compact equation (126) read with equations (128), (132) , (136) and different functions Q(z) , Q(x), R(z ,t) and f(z ,t) outlined for X(z) and Y(z) will determine separate decoupled new expressions for X(z) and Y(z) in the complex z plane cut along (-1,1) as

$$X(z) \, T(z) = f_{Xd}(z,t) \; + z \; \; N(z) \, [\, C \, +$$

$$\int_0^1 \; U(u) \, f_{xd}(u,t) \, d\,u \, / \, (N^+(u) \; T^-(u) \; (\,u - z\,) \; ) \, ] \qquad (139)$$

$$Y(z) \, T(z) = f_{yd}(z,t) \; + z \; \; N(z) [\, D \, +$$

$$\int_0^1 \; U(u) f_{yd}(u,t) \, d\,u \, / \, (N^+(u) \; T^-(u) \; (\,u - z\,) \; ) \; ] \qquad (140)$$

where $\quad f_{Xd}(z,t)$ and $f_{yd}(z,t)$ are given by equations (120) and (121) for f(z, t) when Q(z) = X(z) and Q(z) = Y(z) respectively and constants $A_R$ in equation (132) is denoted to be C when Q(x) = X(x) and to be D when Q(x) = Y(x) respectively and those are to be determined by the conditions at infinity and / or at the zeros of T(z) in the cut plane along (-1,1).

**VIII. Three new functions $N_3(z,t)$, $N_4(z,t)$ and $N_5(z,t)$ in terms of N(z):**

We can set in equations (139) and (140)

$$X_N(z,t) \; = \; \int_0^1 \; U(x) f_{Xd}(x,t) \, d\,x \, / \, (N^+(x) \; T^-(x) \; (\,x - z\,) \; ) \qquad (141)$$

$$Y_N(z,t) \; = \; \int_0^1 \; U(x) f_{Yd}(x,t) \, d\,x \, / \, (N^+(x) \; T^-(x) \; (\,x - z\,) \; ) \qquad (142)$$

We shall use $f_{Xd}(z,t)$ and $f_{Yd}(z,t)$ from equations (120) and (121) to equations (141) and (142) to express $X_N(z,t)$ and $Y_N(z,t)$ as follows:

$$X_N(z,t) = (\, \tfrac{1}{2} \, - B \,) \, N_4(z,t) + N_5(z,t) \qquad (143)$$

$$Y_N(z,t) = (\, 1 - A \,) \; N_4(z,t) \; + \; N_3(z,t) \, / \, N(0) \, . \qquad (144)$$

where

$$N_3(z,t) = \int_0^1 \; U(y) N(-y) \exp(-t/y) \, d\,y \, / \, (N^+(y) \; T^-(y) \; (\,y - z\,) \; ) \qquad (145)$$

$$N_4(z,t) = \int_0^1 y\, U(y) N(-y) \exp(-t/y)\, d\,y\, /\, (N^+(y)\, T^-(y)\, (y-z))\quad (146)$$

$$N_5(z,t) = \int_0^1 y\, U(y) N(-y) \exp(-t/y) V_{20}(-y,t) d\,y\, /(N^+(y) T^-(y)(y-z))\ (147)$$

We shall now express $N_3(z,t)$ , $N_4(z,t)$ and $N_5(z,t)$ in terms of $N(z)$ . In doing so we have to determine the behaviour of the following functions when $z \to \infty$:

$$L_3(z,t) = \exp(-t/z)\, N(-z)\, /\, (z\, N(z))\qquad\qquad (148)$$

$$L_4(z,t) = \exp(-t/z)\, N(-z)\, /\, N(z)\qquad\qquad (149)$$

$$L_5(z,t) = \exp(-t/z)\, N(-z)\, V_{20}(-z,t)\, /\, N(z)\qquad (150)$$

We shall state the properties of $L_3(z,t)$ , $L_4(z,t)$ and $L_5(z,t)$ as follows:

$L_3(z,t)$ is analytic in the complex z plane with Re z >0 and cut along (0,1), tends to 0 as z→0+ , O(1/z) as z→∞ , essential singularity at z→0-, removable singularity at z=1.
$L_4(z,t)$ is analytic in the complex z plane with Re z >0 and cut along (0,1), tends to 0 as z→0+ , -1+O(1/z) as z→∞ , essential singularity at z→0-, removable singularity at z=1.
$L_5(z,t)$ is analytic in the complex z plane with Re z >0 and cut along (0,1), tends to 0 as z→0+ , O(1/z) as z→∞ , essential singularity at z→0-, removable singularity at z=1 where $V_{20}(-z,t)$ is analytic in the complex z plane with Re z >0 , tends to 0 as z→0+ , O(1/z) as z→∞ and essential singularity at z→0-.

We shall consider the same contour for defining contour integral for $N_2(z,t)$ in article (V) herein to determine $N_3(z,t)$ , $N_4(z,t)$ and $N_5(z,t)$ in terns of $N(z)$ as follows :

$$L_n(z,t) = (2\pi i)^{-1} \int_{C_1} L_n(w,t)\, d\,w\, /\, (w-z)\, )\ ,\ n=3,4\ \text{and}\ 5.\quad (151)$$

where $C_1 = L_1\, U\, L_2$ . Here contour $L_1$ is a sufficiently large in the complex w plane to contain the cut along (0,1) within and $L_2$ is a semicircle with center at the origin and of very large radius R in Re z >0 to avoid the essential singularity at z →0- with diameter on the imaginary axis and $L_1$ lies interior to this $L_2$ and z lies outside $L_1$ but inside $L_2$ . Both $L_1$ and $L_2$ are taken in counter clockwise sense. The functions $N(-w)$ , $1/N(w)$ , $V_{20}(-z,t)$, and $\exp(-t/z)$ are analytic and nonzero in the annulus $C_1$ . We shall now apply the Cauchy Integral theorem on the functions

$$L_3(w,t) = \exp(-t/w)\, N(-w)\, /\, (w\, N(w))\qquad\qquad (152)$$

$$L_4(w, t) = \exp(-t/w) \, N(-w) \, / \, N(w) \qquad (153)$$

$$L_5(w, t) = \exp(-t/w) \, N(-w) \, V_{20}(-w, t) \, / \, N(w) \qquad (154)$$

to obtain, using the same procedure as in the case $N_2(z, t)$, for $n = 3, 4$ and $5$,

$$L_n(z, t) = -N_n(z, t) + (2\pi i)^{-1} \int_{C1} L_n(w, t) d\,w / (w-z)), \qquad (155)$$

where

$$(2\pi i)^{-1} \int_{C1} L_3(w, t) d\,w / (w-z)) = -(2\pi i)^{-1} \int_{-i\infty}^{+i\infty} L_3(w, t) d\,w / (w-z)),$$

$$= -V_{30}(z, t) \qquad (156)$$

$$(2\pi i)^{-1} \int_{C1} L_4(w, t) d\,w / (w-z)) = -(2\pi i)^{-1} \int_{-i\infty}^{+i\infty} L_4(w, t) d\,w / (w-z)) \; -1/2$$

$$= -V_{40}(z, t) \; - 1/2 \qquad (157)$$

$$(2\pi i)^{-1} \int_{C1} L_5(w, t) d\,w / (w-z)) = -(2\pi i)^{-1} \int_{-i\infty}^{+i\infty} L_5(w, t) d\,w / (w-z)),$$

$$= -V_{520}(z, t) \qquad (158)$$

Hence from equations (155), using equations (156-158), we get

$$N_3(z, t) = -\exp(-t/z) N(-z) / (z \, N(z)) - V_{30}(z, t), \qquad (159)$$

$$N_4(z, t) = -\exp(-t/z) N(-z) / N(z) - V_{40}(z, t) \; - 1/2, \qquad (160)$$

$$N_5(z, t) = -\exp(-t/z) N(-z) V_{20}(-z, t) / N(z) - V_{520}(z, t), \qquad (161)$$

The equations (159-161) will give the three unknown functions $N_3(z, t)$, $N_4(z, t)$ and $N_5(z, t)$ and some new convergent integrals $V_{30}(z, t)$, $V_{40}(z, t)$ and $V_{520}(z, t)$ in terms of known $N(z)$.

### IX. Explicit forms of X(z) and Y(z) functions:

We shall now substitute the results of equations (159), (160) and (161) in equations (139) and (140), after simplification to get explicit forms of $X(z)$ and $Y(z)$, in terms of $H(z)$, $N(-z)$ as follows :

$X_N(z, t) = -( \frac{1}{2} - B ) [ \exp(-t/z) N(-z) / N(z) + V_{40}(z, t) + 1/2 ]$

$-[ \exp(-t/z) N(-z) V_{20}(-z, t) / N(z) + V_{520}(z, t) ]$     (162)

$Y_N(z, t) = -( 1 - A ) [ \exp(-t/z) N(-z) / N(z) + V_{40}(z, t) + 1/2 ]$

$- [ \exp(-t/z) N(-z) / ( z N(z) ) + V_{30}(z, t) ] / N(0)$ .     (163)

With equations (162) and (163) , equations (139) , (140) after simplifications now take explicitly new forms for $X(z)$ and $Y(z)$ in terms of $N(z)$ as follows for $Re(z) > 0$ :

$X(z) T(z) = z N(z) [ C - [1/2 + V_{40}(z, t) ][1/2 - B] - V_{520}(z, t) ]$     (164)

$Y(z) T(z) = z N(z) [ D - [1/2 + V_{40}(z, t) ][1 - A] - V_{30}(z, t) / N(0) ]$ (165)

We know following Das [ 39] , Das[40] that when z is outside cut along (-1, 1)

$H(z) T(z) N(0) = N(z) (1 - k z ) ]$                           (166)

Equations (164) and (165) with equation (166) take explicitly new form of $X(z)$ and $Y(z)$ for $Re(z) > 0$ in terms of $H(z)$ and $N(z)$ as

$X(z) = [ z N(0) H(z) / ( 1 - k z ) ]$

$[ C - [1/2 + V_{40}(z, t) ][1/2 - B] - V_{520}(z, t) ]$ .     (167)

$Y(z) = [ z N(0) H(z) / ( 1 - k z ) ]$

$[ D - [1/2 + V_{40}(z, t) ][1 - A] - V_{30}(z, t) / N(0) ]$     (168)

where $V_{30}(z, t)$ , $V_{40}(z, t)$ and $V_{520}(z, t)$ are given by equations (156) , (157) and (158) read with equations (152), (153) and (154) respectively.

When z is real , z=x and lies in (0,1) , equations (167) and (168) will take the same form replacing z by x .

We shall now substitute the values of equations (141- 144) to equations (139) and (140) to get other explicit forms of $X(z)$ and $Y(z)$ for $Re(z) > 0$ as

$$X(z) = z N(0) H(z) [C + N_4(z , t)[1/2- B] + N_5(z , t)] / (1 - k z)$$

$$+ z \exp(-t/z)N(-z) [1/2- B + V_{20}(-z ,t)] / T(z), \quad (169)$$

$$Y(z) = z N(0) H(z)[ D + [1 - A] N_4(z , t) + N_3(z , t) / N(0) ] / (1-k z)$$

$$+ z \exp(-t/z)N(-z) [1- A + 1 / (z N(0)) ] / T(z) . \quad (170)$$

When z is real , z = x , x lies in (0 ,1) we get from equations (169) and (170) the explicit forms of X(x) and Y(x) as

$$X(x) = x N(0) H(x) [C + N_4(x , t)[1/2- B] + N_5(x , t)] / (1 - kx)$$

$$+ x \exp(-t/x)N(-x) [1/2- B + V_{20}(- ,t)] \ T_c(x) ( T^2_c ( x) + \pi^2 \ x^2 \ U^2(x))^{-1}$$

$$(170)$$

$$Y(x) = x N(0) H(x)[ D + [1 - A] N_4(x , t) + N_3(x , t) / N(0) ] / (1-k x)$$

$$+ x \exp(-t/x)N(-x) [1- A + 1 / (z N(0)) ] \ T_c(x) ( T^2_c ( x) + \pi^2 \ x^2 \ U^2(x))^{-1}$$

$$(171)$$

where $N_3 (x ,t)$ , $N_4(x, t)$ and $N_5(x ,t)$ are taken as Cauchy principal value sense in equations (145) ,(146) and (147) and where

$$1/ [N^+(u) \ T^-(u)] = ( 1- k^2 u^2) N(-u) / [ N^2(0) ( T_0^{\ 2}(u) + \pi^2 u^2 \ U^2(u) ) ]$$
$$(172)$$
$$N(0) = k / D_1^{-1/2} \qquad (173)$$

$$D_1 = [ 1 - 2 U_0] \qquad (174)$$

Equations (139) and (140) can be written as when z in the complex plane cut along (-1,1)

$$X(z) T(z) = f_{xd}(z,t) [ 1 + z N(z) N_1(z) ] + z \ N(z) [C +$$

$$\int_0^1 U (u) [ f_{xd}(u, t) - f_{xd}(z ,t) ] du / [N^+(u) \ T^-(u) ( u - z) ) ] \ , (175)$$

$$Y(z) T(z) = f_{yd}(z,t) [ 1 + z \ N(z) N_1(z) ] + z \ N(z)[ D +$$

$$\int_0^1 U (u) [f_{yd}(u, t) - f_{yd}(z,t) ] \ du / [N^+(u) \ T^-(u) ( u - z) ) ] . \ (176)$$

Using equations (103 , 166) in equations (175) and (176) we get otherexplicit expression of X(z) and Y(z) in the complex z plane R l z >0 as

$X(z) = [H(z) / (1 - k z)] [\{ f_{xd}(z,t) (1 - z N(0)) + z N(0) C \}$

$+ z N(0) \int_0^1 U(u) [f_{xd}(u, t) - f_{xd}(z,t)] du / [N^+(u) T^-(u) (u - z)]]$, (177)

$Y(z) = [H(z) / (1 - k z)] [\{ f_{yd}(z,t) (1 - z N(0)) + z N(0) D \}$

$+ z N(0) \int_0^1 U(u) [f_{yd}(u, t) - f_{yd}(z,t)] du / [N^+(u) T^-(u) (u - z)]]$. (178)

where $f_{xd}(z,t)$ and $f_{yd}(z,t)$ are given by equations (120) and (121).

When $z \to u$ when $0 < u < 1$ equations (177) and (178) gives the otherexplicit forms of $X(u)$ and $Y(u)$ in Cauchy principal value sense as

$X(u) = [H(u) / (1 - k u)] [\{ f_{xd}(u,t) (1 - u N(0)) + u N(0) C \}$

$+ u N(0) P \int_0^1 U(y) [f_{xd}(y, t) - f_{xd}(u,t)] dy / [N^+(y) T^-(y) (y - u)]]$., (179)

$Y(u) = [H(u) / (1 - k u)] [\{ f_{yd}(u,t) (1 - u N(0)) + u N(0) D \}$

$+ u N(0) P \int_0^1 U(y) [f_{yd}(y, t) - f_{yd}(u,t)] dy / [N^+(y) T^-(y) (y - u)]]$, (180)

where $f_{xd}(u,t)$ and $f_{yd}(u,t)$ are given by equations (120) and (121), P before the integrals indicate the value of the integrals in Cauchy principal value sense

Equations (139) and (140) using equation (166) take new another forms in the complex z plane cut along (-1,1) as :

$X(z) = f_{xd}(z,t) / T(z) + z N(0) [H(z)/(1 - k z)]$

$[C + \int_0^1 U(u) f_{xd}(u, t) d u / (N^+(u) T^-(u) (u - z))]$, (181)

$Y(z) = f_{yd}(z,t) / T(z) + z N(o) [H(z) / (1 - k z)]$

$[D + \int_0^1 U(u) f_{yd}(u, t) d u / (N^+(u) T^-(u) (u - z))]$. (182)

When RI (z) > 0 and $z \to u$, $0 < u < 1$ from above and below the cut along (0,1), from equations (181) and (182), we get new another form:

$X(u) = f_{xd}(u,t) T_c(u) / (T_c^2(u) + \pi^2 u^2 U^2(u)) + u N(0) [H(u) / (1 - k u)]$

$$[ C + P \int_0^1 U(y) f_{xd}(y, t) \, dy / (N^+(y) \, T^-(y) \, (y - u) \, ) ] \qquad , (183)$$

$$Y(u) = f_{yd}(u,t) \, T_c(u) / (\, T_c^2(u) + \pi^2 u^2 \, U^2(u) \,) + u \, N(0) \, [H(u) / (\, 1 - k \, u \,) ]$$

$$[ D + P \int_0^1 U(y) f_{yd}(y, t) \, dy / (N^+(y) \, T^-(y) \, (y - u) \, ) ] \qquad . (184)$$

where $f_{xd}(u, t)$ and $f_{yd}(u, t)$ are given by equations (120) and (121) and

$$1 / (\, N^+(u) \, T^-(u)) = (1 - k^2 u^2) \, N(-u) / [\, N^2(0) \, (T_c^2(u) + \pi^2 u^2 \, U^2(u)) \,] \, . (185)$$

### X. Constants A, B, C and D :

In non-conservative cases, the constants A and B will be determined from the constraints (124) and (125) as follows ( those are dependent on X(x) and Y(x) ) :

$$B = k \, [ \, N_2(-1/k, t) \; X_0(1/k) - X_{0N2}(1/k, t) \, ] - k \, N_2(-1/k, t)$$

$$+ k \, [ \, 1 - X_0(1/k) \, ] \exp(k \, t) / N(-1/k) \, , \qquad (186)$$

$$A = k \, [ \, N_2(-1/k, t) \; Y_0(1/k) - Y_{0N2}(1/k, t) \, ] - \; N_1(-1/k)$$

$$+ k \, [ \, 1 - \exp(\, k \, t) \, Y_0(1/k) \, ] / N(-1/k) \, , \qquad (187)$$

where

$$X_0(1/k) = \int_0^1 X(u) \, U(u) \, du / (\, 1 - k \, u \,) \, , \qquad (188)$$

$$X_{0N2}(1/k, t) = \int_0^1 X(u) U(u) \, N_2(-u, t) \, du / (\, 1 - k \, u \,) \, , \qquad (189)$$

$$Y_0(1/k) = \int_0^1 Y(u) \, U(u) \, du / (\, 1 - k \, u \,) \, , \qquad (190)$$

$$Y_{0N2}(1/k, t) = \int_0^1 Y(u) \, U(u) \, N_2(-u, t) \, du / (\, 1 - k \, u \,) \, . \qquad (191)$$

In conservative cases, the constants A and B can be obtained from equations (186) and (187) in the limit of $k \rightarrow 0$ ( those are dependent on X(x) and Y(x) ) as

$$A = [\, 1 - y_0 / 2 + (\, 2 \, U_2)^{-1/2} \, ] \qquad (192)$$

$$B = [\, 1 - x_0 \, ] \, / \, 2 \qquad (193)$$

where

$$x_0 = \int_0^1 X(u) U(u) d u \qquad (194)$$

$$y_0 = \int_0^1 Y(u) U(u) d u \qquad (195)$$

We shall determine the constants A and B not dependent on X(x) and Y(x) from equations (164) and (165) using zeros of T(z) at $z = 1/k$ and $z = -1/k$ as

$$B = N_x(k, t) / D_x(k, t), \qquad (196)$$

$$A = N_y(k, t) / D_x(k, t), \qquad (197)$$

where

$$N_x(k, t) = N_{x1}(k, t) - (k^2/(2 D_1))[N_{x2}(k, t) - N_{x3}(k, t)], (198)$$

$$N_y(k, t) = N_{y1}(k, t) - (k^2/(2 D_1))[N_{y2}(k, t) + N_{y3}(k, t)], (199)$$

$$N_{x1}(k, t) = \int_0^1 [1 - \exp(-t/u) N(-u) N_2(-u, t)] U(u) N(-u) d u / (T_0^2(u)$$

$$+ \pi^2 u^2 U^2(u)), \qquad (200)$$

$$N_{x2}(k, t) = [1/N(-1/k) + 1/N(1/k)], \qquad (201)$$

$$N_{x3}(k, t) = \exp(k t) N(1/k) N_2(1/k, t) / N(-1/k)$$

$$+ \exp(-k t) N(-1/k) N_2(-1/k, t) / N(1/k), \qquad (202)$$

$$N_{y1}(k, t) = \int_0^1 \exp(-t/u)[1 - u N_1(-u) N(-u)] U(u) N(-u) d u / (T_0^2(u)$$

$$+ \pi^2 u^2 U^2(u)), \qquad (203)$$

$$N_{y2}(k, t) = \exp(-k t) / N(1/k) + \exp(k t) / N(-1/k), \qquad (204)$$

$$N_{y3}(k, t) = k^{-1} [\exp(k t) N(1/k) N_1(1/k, t) / N(-1/k)$$

$$- \exp(-k t) N(-1/k) N_1(-1/k, t) / N(1/k)], \qquad (205)$$

$$D_x(k, t) = [\exp(k t) N(1/k) / N(-1/k) - \exp(-k t) N(-1/k) / N(1/k)] k / (2 D_1)$$

$$+ \int_0^1 \exp(-t/u) u N^2(-u) U(u) d u / (T_0^2(u) + \pi^2 u^2 U^2(u)),$$

$$(207)$$

In non conservative cases, C and D (dependent on B and A) will be obtained from equations (139) and (140) as

$$C = \int_0^1 k\, U(u)\, f_{xd}(u, t)\, d\,u\, /\, (N^+(u)\, T^-(u)\, (\,1 - k\,u\,)\,)\,]$$

$$- f_{xd}(1/k, t)\, k\, /\, N(1/k)\,, \qquad (208)$$

$$D = \int_0^1 U(u) f_{yd}(u, t)\, d\,u\, /\, (N^+(u)\, T^-(u)\, (\,1 - k\,u\,)\,)$$

$$- f_{yd}(1/k, t)\, k\, /\, N(1/k)\,. \qquad (209)$$

We can determine C and D from equations (208) and (209) in the limit of k→0 for conservative cases as

$$C = 1 - B \qquad (210a)$$

$$D = 1 - A \qquad (210b)$$

We can determine the constants A , B free from dependence on X(x) and Y(x) for conservative cases in the limit of k→0 from equations(196) and (197) as follows:

$$B = N_x(0, t)\, /\, D_x(0, t)\,, \qquad (210\ c)$$

$$A = N_y(0, t)\, /\, D_x(0, t)\,, \qquad (211\ d)$$

where

$$N_x(0, t) = N_{x1}(0, t)\, - (2U_2)^{-1}[\, N_{x2}(0, t) - N_{x3}(0, t)\,]\, /\, 2\,, \qquad (212)$$

$$N_y(0, t) = N_{y1}(0, t)\, - (2U_2) - 1[\, N_{y2}(0, t) + N_{y3}(0, t)\,]\, /2\,, \qquad (213)$$

$$N_{x1}(0, t) = \int_0^1 [\, 1 - \exp(-t/u)\, N(-u)\, N_2(-u, t)\,]\, U(u)\, N(-u)\, d\,u\, /\, (\,T_0{}^2(u) + \pi^2\, u^2\, U^2(u)\,)\,, \qquad (214)$$

$$N_{x2}(0, t) = 2\,(\,1 - \theta_0\,) \qquad (215)$$

$$N_{x3}(0, t) = [\,-\,t + (\pi)^{-1} \int_0^\infty (\, L_2(-i\,y) + L_2(i\,y)\,)\, d\,y\,] \qquad (216)$$

$$N_{y1}(0, t) = \int_0^1 \exp(-t/u)[1 - u\, N_1(-u)\, N(-u)\,]\, U(u)\, N(-u)\, d\,u\, /\, (\,T_0{}^2(u) + \pi^2\, u^2\, U^2(u))\,, \qquad (217)$$

$$N_{y2}(0, t) = 2\,[\,t + (1 - \theta_0)\,] \qquad (218)$$

$$N_{y3}(0, t) = 2\,[\,(1 - \theta_0) - 1/\, N(0)\,] \qquad (219)$$

$$D_x(0, t) = [\,2\,(\,\theta_0 - 1) - t\,]\,(\,2\, U_2)^{-1}$$
$$+ \int_0^1 \exp(-t/u)\, u\, N^2(-u)\, U(u)\, d\,u\, /\, (\,T_0{}^2(u) + \pi^2\, u^2\, U^2(u))\,,$$

$$\theta_0 = \pi^{-1} \int_0^1 \theta(x)\, dx \, . \tag{220a}$$

(220)

where $\theta(x)$ will be given by equation(43).

If we take $t \to \infty$ to equations(208) , (209) ,( 196 ) and (197)  we get for non conservative cases

$$A = C = 1 - k / N(0) \tag{221a}$$

$$B = D = 0 \tag{221b}$$

and for conservative cases if we take $t \to \infty$  to equations(208) ,(209) , (221a )  and ( 221b) in the limit $k \to 0$

$$A = C = 1 \tag{222a}$$
$$B = D = 0 \tag{222b}$$

We can determine the   constants A , B , C and D   in semi-infinite atmosphere  when $t \to \infty$ in non conservative cases as follows :

$$B = D = 0 \tag{223a}$$

$$A = C = 1 - k / N(0) \, , \tag{223b}$$

## XI. Further forms of X-and Y- functions:

Equations (164) and (165) after substitution of expression of C and D from equations ( 208) and (209) and after minor simplification   give the explicit form of $X(z)$ and $Y(z)$ in the complex plane cut along (-1,1)

$$X(z)\, T(z) = z\, N(z) \; [ [V_{40}(1/k, t) - V_{40}(z ,t)][1/2 - B] - V_{520}(z ,t) ]. \tag{224}$$

$$Y(z)\, T(z) = z\, N(z)[ \, [ V_{40}(1/k,t) - V_{40}(z, t) \,][1 - A] - V_{30}(z , t) \, / \, N(0) \, ] \tag{225}$$

Using equation (166)  to equations ( 224) and (225)  we get another explicit form of $X(z)$ and $Y(z)$ in complex $Rl\, z > 0$  ,

$$X(z) = [z\, N(0)\, H(z) / ( 1 - k\, z ) ]$$

$$[ \, [V_{40}(1/k, t) - V_{40}(z ,t)][1/2 - B] - V_{520}(z ,t) \, ] \tag{226}$$

$$Y(z) = z\, N(0)\, H(z) / ( 1 - k\, z )\; ]$$

$$[ \, [ V_{40}(1/k,t) - V_{40}(z, t) \,][1 - A] - V_{30}(z , t) \, / \, N(0) \, ] \tag{227}$$

where $V_{30}(z,t)$, $V_{40}(z,t)$, $V_{520}(z,t)$ are given by equations (156), (157). and (158).

When $z \rightarrow u$, $0 < u < 1$, equations (226) and (227) give the explicit expression as

$X(u) = [u\, N(0)\, H(u)\, /\, (1-k\,u)]$

$[\,[V_{40}(1/k,t) - V_{40}(u,t)][1/2 - B] - V_{520}(u,t)\,]$ \hfill (228)

$Y(u) = u\, N(0)\, H(u) / (1-k\,u)]$

$[\,[V_{40}(1/k,t) - V_{40}(u,t)][1-A] - V_{30}(u,t)\,/\,N(0)\,]$ \hfill (229)

where $V_{30}(u,t)$, $V_{40}(u,t)$ and $V_{520}(u,t)$ are to be used as integrals in Cauchy principal value sense.

where $V_{30}(z,t)$, $V_{40}(z,t)$, $V_{520}(z,t)$ are given by equations (156), (157) and (158).

## XII. H- functions:

If we take $t \rightarrow \infty$ in equations (120), (121), (177), (178) we see that for z out side the cut along (-1,1) in complex plane,

$X(z) = H(z)\,[1 - z\,N(0) + C\,z\,N(0)]\,/\,(1-k\,z) = H(z)$, \hfill (230)

$.Y(z) = 0$. \hfill (231)

If we take $t \rightarrow \infty$ in equations (120), (121), (179), (180) we see that for z=x x is real and lie in (0,1)

$X(x) = H(x)\,[1 - x\,N(0) + C\,x\,N(0)]\,/\,(1-k\,x) = H(x)$, \hfill (232)

$Y(x) = 0$.

\hfill (233)

New forms of H- functions are in Das [39 ,40]

**Conclusion:** The explicit forms for X- and Y- functions so far determined in the literature are dependent on the X- and Y- functions or on some other unknown function satisfying singular integral equations. In this paper the X- and Y- functions are decoupled from coupled Riemann Hilbert problem in a new manner using the theory of linear singular integral equations and of contour integration in terms of two new functions which are expressed in terms of a known N- function in the complex plane. New decoupled forms of linear non homogeneous integral equations for each of X- and Y- functions in complex plane are determined . Determination of two new decoupled linear non homogeneous Riemann Hilbert problems using the theory of linear singular integral equations and contour integration is completely  new in the literature of X- and Y- functions . Our solution from new linear decoupled integral equations for each of X- and Y- function are also completely new  and dependent only on five unknown functions and on one known N- function .All those five unknown functions are represented in terms of known N-function . Hence our two forms of expression for each X- and Y- function  is dependent on a known N-function only . The convergence of the integrals over cut along imaginary axis in the complex plane and numerical evaluation of those X- and Y- functions  with asymptotic expansions from those new expressions are prepared and  awaiting communication  .

**Acknowledgement :** I express my heartfelt gratitude to the Library and support personnel of Saha Institute of Nuclear Physics , Salt Lake , Kolkata , West Bengal , India  for their extended whole hearted support  and  the Department of Mathematics , Heritage Institute  of Technology  , Anandpur , Kolkata  for their constant encouragement .

……………………………………………………………………………..